\documentclass[preprint,12pt]{elsarticle}

\usepackage[a4paper, left=1in, right=1in, top=0.5in, bottom=0.5in]{geometry}
\usepackage{graphicx,natbib,fancyhdr} 
\setcitestyle{authoryear,open={(},close={)}}
\usepackage{longtable,tabularx,rotating,multirow,url}
\usepackage{amsmath,comment,subcaption}
\usepackage{amsfonts,color,gensymb}
\usepackage{amssymb}
\usepackage{verbatim}
\usepackage{lineno}
\usepackage{draftwatermark}
\SetWatermarkText{Preprint}
\SetWatermarkScale{1.0}
\SetWatermarkLightness{0.85}
\begin{document}

\begin{frontmatter}



\title{Auto-Regressive Standard Precipitation Index: A Bayesian Approach for Drought Characterization}




\author[IITI]{Soham Ghosh}

\affiliation[IITI]{organization={Department of Economics, School of Humanities and Social Sciences},
            addressline={Indian Institute of Technology Indore}, 
            city={Indore},
            postcode={453552}, 
            state={Madhya Pradesh},
            country={India}}

\author[IIMI]{Sujay Mukhoti}

\affiliation[IIMI]{organization={Operations Management and Quantitative Techniques Area},             
            addressline={Indian Institute of Management Indore}, 
            city={Indore},
            postcode={453556}, 
            state={Madhya Pradesh},
            country={India}}

\author[IITI]{Pritee Sharma}


\begin{abstract}
    This study proposes Auto-Regressive Standardized Precipitation Index (ARSPI) as a novel alternative to the traditional Standardized Precipitation Index (SPI) for measuring drought by relaxing the assumption of independent and identical rainfall distribution over time. ARSPI utilizes an auto-regressive framework to tackle the auto-correlated characteristics of precipitation, providing a more precise depiction of drought dynamics. The proposed model integrates a spike-and-slab log-normal distribution for zero rainfall seasons. The Bayesian Monte Carlo Markov Chain (MCMC) approach simplifies the SPI computation using the non-parametric predictive density estimation of total rainfall across various time windows from simulated samples. The MCMC simulations further ensure robust estimation of severity, duration, peak and return period with greater precision. This study also provides a comparison between the performances of ARSPI and SPI using the precipitation data from the Colorado River Basin (1893–1991). ARSPI emerges to be more efficient than the benchmark SPI in terms of model fit. ARSPI shows enhanced sensitivity to climatic extremes, making it a valuable tool for hydrological research and water resource management. 
\end{abstract}








\end{frontmatter}

%


\section{Introduction} \label{sec : Intro}

Drought is recognized as a significant natural disaster that affects every region of the world \citep{wilhite2007understanding}. Its effects are felt widely across agriculture, socio-economics, and the environment. Existing drought indicators assume that the precipitation data are independent over time, although there is substantial amount of literature indicating existence of weak-stationarity in precipitation data. In this work, we develop a drought index accounting for the weak stationarity using a Bayesian auto-regressive model and study the key characteristics of drought based on the proposed index.

\par A key challenge in studying drought is the lack of a universally accepted definition, as experts from various disciplines define it based on their specific perspectives. Accordingly, drought is categorized into four types: Meteorological, Agricultural, Hydrological, and Socio-Economic \citep{yihdego2017highlighting}. Meteorological drought refers to a deficiency in precipitation over a given time period. Agricultural drought is linked to prolonged insufficient crop yields and inadequate soil moisture. Socio-Economic drought occurs when water resource systems are unable to meet societal water needs, while hydrological drought is characterized by a lack of surface and subsurface water resources. Efforts to refine drought definitions continue, focusing on its impacts on different societal and environmental aspects. 
\par Numerous drought indices have been developed so far to monitor and assess various types of drought. One of the most popular drought indices is the Standardized Precipitation Index (SPI), which is used for measuring meteorological drought across the globe \citep{ntale2003drought,tsakiris2007regional,zhao2020frequency,yildirim2022spatiotemporal,ullah2023spatiotemporal}. SPI provides valuable insights including a spectrum from arid to drought conditions, which can be challenging to assess otherwise \citep{mukhawana2024hydrological}. The World Meteorological Organization (WMO) recommends SPI as a primary indicator of drought \citep{monitoring2006concepts}.

\par The computation of the Standardized Precipitation Index (SPI) is based on the premise that the moving totals of rainfall (MTR) data over a window of time are independent and identically distributed (i.i.d.) realizations of Gamma random variables \citep{mckee1993relationship}. Numerous studies have shown that rainfall series from various regions and time windows tend to exhibit positive skewness \citep{saha2018disparity, haktanir2010frequency, sankarasubramanian1999investigation, yu1994estimating}. The use of the Gamma distribution helps to address the non-negativity as well as the skewness in precipitation data. However, the Gamma distribution also presents certain limitations. For instance, it has a thin tail, meaning it underestimates the likelihood of extreme rainfall events, making it less suitable for modeling rainfall series that include frequent heavy rainfalls. Additionally, the Gamma distribution may not fit well with data that have extremely low variance, potentially leading to poor parameter estimation due to over-dispersion. 

\par A key challenge in constructing the SPI, which has not been thoroughly addressed in the literature, is how to model MTR data that do not satisfy the i.i.d. assumption. The i.i.d assumption reflects strict stationarity of the MTR series, which in many real-world situations, including precipitation series, is often violated. \cite{das2020non} developed a drought index for non-stationary precipitation data and demonstrated its efficiency in capturing different drought characteristics using a generalized additive model. However, weak-stationarity (or time independent auto-correlation) in precipitation data is most commonly reported in the literature. For example, \cite{soltani2007use} reported significant variation in the temporal patterns of monthly rainfall series across much of Iran. \cite{obeysekera1987parameter} also observed auto-correlation in hourly rainfall data in northeastern Colorado, while \cite{delleur1978stochastic} demonstrated similar findings for monthly rainfall series in 15 river basins in Indiana, Kentucky, and Illinois. Additionally, \cite{labadie1981worth} identified significant short-term auto-correlation in rainfall data for northern San Francisco, and \cite{mahmud2017monthly} documented auto-correlation in monthly rainfall series with a 12-month lead time across 30 meteorological stations in Bangladesh.

\par Another notable limitation of the Gamma distribution is its complexity in modeling auto-correlated MTR data. In contrast, the log-normal distribution distribution has been recommended as a viable alternative \citep[see][for example]{angelidis2012computation, lloyd2002drought,lana2000some,stagge2015candidate,shoji2006statistical,moccia2022spi}. It provides a more parsimonious model to accommodate non-negative support, positive skewness and auto-correlation. The conditional mean of log-transformed MTR data can be effectively represented as a stationary Gaussian auto-regressive (AR) process under the assumption of log-normality \citep{wilks2011statistical}. A key challenge in constructing the Standardized Precipitation Index (SPI) is accurately accounting for the occurrence of zero-rainfall days, which vary with seasonal changes. Addressing this zero-inflation alongside auto-correlation proves to be difficult. In this paper, we propose a spike-and-slab log-Normal AR model for MTR data, where MTR is modeled in two parts: a probability spike at zero and the remainder distributed as a log-Normal variable with AR conditional mean \citep[see][for review of spike and slab]{lempers1971posterior,mitchell1988bayesian}. During the construction of the Standardized Precipitation Index (SPI), it's necessary to assess the cumulative distribution function (CDF) at the de-seasonalised MTR values and create a Gaussian mapping based on the resulting CDFs. It's worth noting that parameter and CDF estimation in such a complex model can be challenging. In this paper we use Bayesian method for parameter estimation and simulate predictive densities of MTR for CDF estimation.

\par Recent advances in computational techniques, such as Markov Chain Monte Carlo (MCMC) methods, have contributed to the widespread adoption of Bayesian inference in hydrology \citep{TANG2016551,SMITH201529}. The Bayesian approach incorporates uncertainty in model parameters by assigning known prior probability distributions. Bayesian inference seeks to update these priors by deriving the posterior distributions of the parameters based on observed data. However, deriving the exact posterior can be challenging for complex models. This challenge is typically addressed by using MCMC methods, like the Gibbs sampler or Metropolis-Hastings algorithms, to generate samples from the posterior distribution (see \cite{chib1995understanding}, for a review). In practice, the posterior mean is often reported as the Bayesian estimate, aligning with the frequentist perspective.

\par In this study, the proposed spike-and-slab log-Normal AR model is structured as a Bayesian network. Here, the nodes represent MTR distributions, which are either degenerate at zero on dry days or log-Normal with an auto-regressive (AR) mean function on rainy days. The prior for these two states is encoded by the edges, modeled as a Bernoulli distribution with the probability of success representing the likelihood of no rainfall. The weak stationarity of the MTR distribution is accounted for by including a Uniform prior on the lagged log-MTR coefficient within the mean function of the log-transformed MTR data, referred to as a log-stationary process. Other parameters are assigned non-informative priors due to a lack of particular prior knowledge.

\par The model's predictive density is obtained by averaging the model's density over the posterior distribution of the parameters. In this Bayesian framework, simulated MTR approximates the posterior predictive distribution, allowing for the empirical estimation of the cumulative distribution function (CDF) at observed precipitation values. Furthermore, uncertainty in drought characteristics, such as severity, duration, and peak, can be assessed using independent and identically distributed (i.i.d.) predictive samples.



\par SPI was invented by a group of scientists at Colorado state University and applied as an experimental tool for drought monitoring in Colorado river basin during its inception. Climate condition of the basin is extremely vulnerable and it faces dual challenges of extreme drought and periodic flooding conditions. In this paper we use the same data, \emph{viz.} the monthly precipitation dataset for the time period 1893--1991, for calculating and comparing the performances of the SPI and the proposed one.

\par The rest of the paper is organized as follows: Section~\ref{sec : ARSPI} introduces the development of ARSPI. The index accounts for the auto-correlated structure of the precipitation time series. Section~\ref{sec: Bayes_est} elaborates a Bayesian MCMC approach for estimating the probabilistic structure of rainfall, leveraging Bayesian methods to impose necessary constraints. Section~\ref{sec : realdata} demonstrates the real application of ARSPI using precipitation data from the Colorado River Basin (1893-1991). Finally, the paper concludes with a discussion of the key findings and their implications.


\section{Precipitation Index for Auto-Correlated Rainfall Data} \label{sec : ARSPI}
\subsection{Conventional SPI}
The Standardized Precipitation Index (SPI), introduced by \cite{mckee1993relationship}, is based on monthly rainfall data. The construction approach can be described as follows: let \( X_1, X_2, \ldots, X_T \) represent i.i.d. monthly rainfall observations collected over \( T \) consecutive months (where \( T \geq 360 \), as per \cite{mckee1993relationship}). A moving total over a window of \( \zeta \) months is computed as 
\[
r_t = \sum_{i=t-\zeta+1}^t X_i, \quad \zeta = 1, 3, 6, \ldots, 72.
\]
The resulting series \( r_t \) is referred to as moving total rainfall (MTR) from here onward. The non-zero MTR, denoted as 
\[
Y_t = I(r_t > 0)r_t, 
\]
where \( I(\cdot) \) is the indicator function, is then fitted to a two-parameter Gamma distribution. The probability density function (pdf) of the Gamma distribution with shape parameter $k$ and scale parameter $\tau$ is given as follows:
\begin{equation}
f_{k,\tau}(y) = \frac{1}{\Gamma(k)\tau^k} y^{k-1} e^{-\frac{y}{\tau}}, \quad y,k,\tau > 0,
\label{Gammapdf}
\end{equation}
where \(\Gamma(k)\) is the Gamma function, defined as \(\Gamma(k) = \int_{0}^{\infty} u^{k-1} e^{-u} \, du\). 
The shape and scale parameters are estimated using the method of moments. The estimates are given by the following equations:
\[
\hat{\alpha} = \frac{1}{4B} \left(1 + \sqrt{1 + \frac{4B}{3}} \right), \quad \hat{k} = \frac{\bar{y}}{\hat{\tau}},
\]
where \( B = \ln(\bar{y}) - \frac{\sum_{j=1}^{m} \ln(y_j)}{m} \), \( y_1, y_2, \ldots, y_m \) are samples from the MTR series, and \( \bar{y} \) is the sample mean. 
Using these estimates, the Cumulative Density Function (CDF) is computed for each observation \( y_1, y_2, \ldots, y_m \) as:
\begin{equation}
\hat{F}(y_t) = \int_{0}^{y_t} f_{k,\tau}(x) \, dx \bigg|_{k=\hat{k}, \tau=\hat{\tau}}, \quad t = 1, 2, \ldots, m,
\label{GamCDF}
\end{equation}
where \( f_{k,\tau}(x) \) represents the Gamma probability density function with shape parameter \( k \) and scale parameter \( \tau \).
The probability of a zero-rainfall window is estimated as 
\[
\pi = \frac{v}{T},
\]
where \( v \) represents the number of zero MTR values, and \( T \) is the total number of observations. Thus the modified CDF of the MTR data is given by:
\begin{equation}
G(y_t) = \begin{cases} 
      \pi, & \text{if } y_t = 0 \\ 
      \pi + (1 - \pi) F(y_t), & \text{if } y_t > 0, \quad t = 1, 2, \ldots, m
   \end{cases}
\label{GamnewCDF}
\end{equation}
Finally, the CDF values are transformed into a standard Normal variate via the inverse Gaussian transformation, \(SPI_t = \Phi^{-1}(G(y_t)),~t=1,2,\ldots m\).
\subsection{Precipitation Index for Auto-Correlated Rainfall}
\cite{lloyd2002drought} demonstrated the effectiveness of the Standardized Precipitation Index (SPI) for drought classification following normalization. Table~\ref{tab : SPItable} presents the index values corresponding to both wet and drought conditions. The SPI is calculated using a moving window across various time scales. While shorter time scales tend to produce frequent fluctuations around zero, longer time scales yield a smoother SPI response to changes in rainfall. However, in regions with low seasonal precipitation, shorter time scales can inaccurately indicate severe drought through highly negative SPI values.

\par In this study, we relax the traditional assumption of independent and identically distributed (i.i.d.) observations commonly used in SPI computation. We introduce a more realistic framework for calculating the Standardized Precipitation Index (SPI) by modeling the logarithm of the mean total rainfall (MTR) during rainy periods, \(\log y_t\), as a weakly stationary Gaussian auto-regressive (AR) process, resulting in a log-normal distribution for non-zero MTR values. 
To account for seasons with zero rainfall, we represent the MTR as a spike-and-slab process with a time-varying probability of zero rainfall (\(r_t = 0\)) within a given window. The MTR at time \(t\) (\(r_t\)) is modeled as follows:
\begin{eqnarray}
    Z_t & \sim & \text{Bernoulli}(\pi_t), \quad 0 < \pi_t < 1 \nonumber \\ 
    r_t & = & Z_t + (1 - Z_t) y_t \nonumber \\
    y_t \mid r_{t-1} & \sim & \text{log-Normal}(\mu_t, \sigma_t^2) \nonumber \\
    logit(\pi_t)&=& \alpha+\phi~logit(\pi_{t-1}) \nonumber \\
    \mu_t & = & e^{\beta_1 + \frac{\sigma_t^2}{2}} r_{t-1}^{\beta_2} \nonumber \\
    \sigma_t & = & \sigma (1 - Z_t), \quad t = 1, 2, \ldots, T
    \label{eqn:ARSPIt}
\end{eqnarray} 
This approach allows for a more general and data-driven representation of rainfall dynamics, particularly in regions where dry periods are frequent.

\par The proposed model suggests that, following a dry period, rainfall events are driven by a log-normal shock with a constant variance $\sigma^2$. During wet periods, the MTR at each time point depends on the preceding rainfall amount. To ensure the expected MTR remains stable and does not diverge, we impose the condition $|\beta_2| < 1$, thereby guaranteeing that $\log \mu_t = \log \mathbb{E}[Y_t \mid r_{t-1}]$ is weakly stationary. The Auto-Regressive Standardized Precipitation Index (ARSPI) at time point $t$ is obtained by applying the inverse Gaussian transformation to $F(r_t) = P[r_t \leq r_t]$, similar to the conventional SPI, ensuring consistent interpretation. However, calculating $F(r_t)$ based on the proposed model is inherently complex, in addition to the challenges involved in estimating the unknown model parameters.
\section{Bayesian Estimation of ARSPI} \label{sec: Bayes_est}
{We employ a Bayesian Monte Carlo Markov Chain (MCMC) approach for estimation \citep{Carlin2008bayesian}. The Bayesian estimation method offers several advantages in this context. First, Bayesian methods facilitate the straightforward imposition of constraints required for ensuring the stationarity of auto-correlated rainfall and the time-dependent rainfall probabilities through the selection of appropriate priors, a task that can be more complex in a frequentist framework.
Another advantage of the Bayesian approach is that it simplifies the estimation process, reducing the computational complexity compared to frequentist methods.

\par In the initial stage of Bayesian analysis, the probabilistic model underlying the MTR data is defined through the likelihood function (joint density) \( L(\mathbf{r} \mid \boldsymbol{\theta}) \), where \( \boldsymbol{\theta} = (\alpha, \phi, \beta_1, \beta_2, \sigma) \in \Theta \). In the Bayesian framework, \( \boldsymbol{\theta} \) is treated as a random vector with associated probability distribution \( \pi(\boldsymbol{\theta}) \), referred to as the prior distribution.
Priors can be subjective, derived from expert knowledge relevant to the data context. Alternatively, they may be non-informative, conveying minimal prior information, and are often considered objective in nature. The objective of Bayesian inference is to determine the posterior distribution of \( \boldsymbol{\theta} \) given the observed data \( \mathbf{r} \), which is obtained using Bayes' theorem:
\begin{equation}
\pi(\boldsymbol{\theta} \mid \mathbf{r}) = \frac{\pi(\boldsymbol{\theta}) L(\mathbf{r} \mid \boldsymbol{\theta})}{\int_\Theta \pi(\boldsymbol{\theta}) L(\mathbf{r} \mid \boldsymbol{\theta}) d\boldsymbol{\theta}}. \label{PosteriorGen}
\end{equation}
Bayesian estimates of the model parameters are typically represented by the posterior mean. For the computation of the Standardized Precipitation Index (SPI), it is necessary to evaluate the cumulative distribution function (CDF) at specified MTR values. In the Bayesian framework, this computation is simplified by deriving the empirical CDF from simulated observations of the predictive distribution of the MTR data. The predictive density of MTR at $i^{th}$ time point (\( \Tilde{r}_i \)) is defined as:
\[
f(\Tilde{r}_i \mid \boldsymbol{r}) = \int_\Theta f(\Tilde{r}_i \mid \boldsymbol{r}, \boldsymbol{\theta}) \pi(\boldsymbol{\theta} \mid \boldsymbol{r}) d\boldsymbol{\theta}
\]
where \( \Theta \) denotes the parameter space. The cumulative distribution function (CDF) \( F(r_t) = P[r \leq r_t \mid \boldsymbol{r}] \) is estimated empirically from a large sample \(\{\Tilde{r}_t^{(i)}, i = 1, 2, \ldots, M\}\) drawn from the predictive density \( f(\Tilde{r}_t \mid \boldsymbol{r}) \). The empirical CDF is given by:
\[
\hat{F}(r_t) = \frac{1}{M} \sum_{i=1}^M I(\Tilde{r}_t^{(i)} \leq r_t)
\]
where \( I(\cdot) \) is the indicator function. Finally, the auto-regressive SPI (ARSPI) is obtained by applying the inverse Gaussian transformation \( \Phi^{-1}\left(\hat{F}(r_t)\right) \). This newly derived index have the interpretation as outlined in Table~\ref{tab : SPItable}. 
\subsection{MCMC Convergence}
Evaluating the convergence of MCMC chains is a critical step in Bayesian analysis. To assess convergence, we employ graphical approaches. The trace plots, generated after discarding the burn-in period and applying thinning (or selecting every ``thin"-$^{th}$ value), visualize the sampled values and provide insights into the behavior of the MCMC sequence. Evidence of non-convergence would be indicated by discernible trends or persistent correlations in the trace plots.

\subsection{Model Adequacy Measure}  
{To assess the fit of the model, we use the Deviance Information Criterion (DIC) as introduced by \cite{spiegelhalter2002bayesian}. The DIC consists of two main components: the deviance (\(\bar{D}\)) and the model complexity (\(p_D\)). The deviance is defined as:
\begin{equation}
    \bar{D} = E_{\boldsymbol{\theta} \mid \mathbf{y}}\left[-2 \log f(\mathbf{y} \mid \boldsymbol{\theta})\right]
\end{equation}
and serves as a measure of how well the model fits the observed data. The complexity of the model (\(p_D\)) is determined by the effective number of parameters and is given by:
\begin{equation}
    p_D = \bar{D} - D(\bar{\boldsymbol{\theta}}) = \bar{D} + 2 \log f(\mathbf{y} \mid \bar{\boldsymbol{\theta}})
\end{equation}
where \(\bar{\boldsymbol{\theta}} = E[\boldsymbol{\theta} \mid \mathbf{y}]\) denotes the posterior mean of the model parameters.}
\section{Colorado Basin Rainfall Data Analysis} \label{sec : realdata}
The rainfall data comprises monthly measurements from 1893 to 1991 in the Colorado River basin. Below, we briefly highlight the significance of this dataset. The Colorado River stretches approximately 2,300 kilometers, originating in the Rocky Mountains of Colorado and Wyoming and flowing to the Gulf of California. The geographical coordinates of the basin range from 31° N to 42° N and 105.5° W to 118° W (see Figure~\ref{fig : colbasin} \citep{Col2012image}). Water resources within the basin are shared among the growing populations of seven U.S. states and parts of Mexico. The basin is home to two of the largest reservoirs, Lake Mead and Lake Powell. In recent decades, the Colorado River Basin has faced substantial drought conditions, marking one of the driest periods in its history. This prolonged drought is exacerbated by rising temperatures and diminishing snowpacks, which are critical for water supply in the region. The ongoing drought has the potential to significantly impact agricultural practices and energy production for a large portion of the population. 

\subsection{Time-series Statistics of the Colorado Rainfall Data}
Figure~\ref{fig : raincol} illustrates the monthly precipitation series for Colorado River basin. The series demonstrates the highly variable nature of monthly rainfall in Colorado, with values ranging from 0 to 9.27 inches. The precipitation series in the basin has fluctuated without any long-term trend over nearly a 100-year period. Auto Correlation Function (ACF) and Partial Auto Correlation Function (PACF) are widely used statistical techniques to determine the relationship between observations in a time series with one another. Plotting the ACF and PACF values against subsequent time lags is a statistical method for choosing appropriate time series model. Figure~\ref{fig : acfplt} depicts ACF plots across different accumulation windows (3,6,12,24 months). The ACF values across all windows gradually decrease with increasing lags and eventually becoming statistically insignificant after some point. Significant spikes in the initial lags indicate short-time dependencies in the data. Lower-order auto-regressive processes are suitable for modeling these gradual decaying patterns in the data. The PACF plots across all accumulation windows reflect a similar pattern (see Figure~\ref{fig : pacfplt}). We observe presence of significant spikes at lag 1 and spikes at subsequent lags dropping quickly. This pattern in the time series across all windows indicates that the Auto-regressive process of order 1 (AR(1)) is the best model. 


To illustrate the practical application of ARSPI using real-world data, we utilize the rainfall dataset from \cite{mckee1993relationship} and apply the proposed index. Additionally, we compare the performance of ARSPI with two other competing models, as outlined below.

\par \textbf{Model: ARSPI with $\pi_t$ }
\par The ARSPI model, as defined in Equation~\ref{eqn:ARSPIt}, is utilized in this analysis. We adopt a hierarchical Bayesian framework for the prior distributions, which are specified as follows:
\begin{eqnarray*}
    \beta_1 &\sim & N(0, \sigma_\beta^2), \\
    \beta_2 &\sim & \text{Uniform}(-1, 1), \\
    \alpha &\sim & N(0, 0.25), \\
    \phi &\sim & \text{Uniform}(-1, 1), \\
    \sigma & \sim& \text{Inverse-Gamma}(\nu/2, \nu/2)\\
    \nu &\sim& \text{Exponential}(0.1)\\
    \sigma_\beta &\sim & \text{Inverse-Gamma}(\nu_1, \nu_2), \\
    \nu_1 &\sim & \text{Exponential}(0.1), \\
    \nu_2 &\sim & \text{Exponential}(0.1).
\end{eqnarray*}

\subsection{MCMC Results}
\par We conducted an MCMC simulation with a total of 450,000 iterations on a dedicated 11th Gen Intel\textregistered Core$^{TM}$ i5-1135G7 processor 8.00~GB RAM with 2667 MHz RAM speed. This involved running three MCMC chains, each consisting of 150,000 iterations, with a thinning interval of 10 and a burn-in period of 5,000, using Just Another Gibbs Sampler (JAGS-Parallel) within R.




\par \textbf{Outcomes:} The convergence of the Markov Chain Monte Carlo (MCMC) chains is verified through trace plots shown in Figure~\ref{fig : spiar1_hbpt3_tr} to \ref{fig : spiar1_hbpt24_tr}. The Deviance Information Criterion (DIC) for this model is presented in Table~\ref{tab : DICtable}. Initially, the DIC decreases with increasing MTRs, but a significant increase is observed as the MTR rises from 12 to 24. 
\par \textbf{Parameter Estimates:} Table~\ref{tab:parest} presents the Bayesian estimates and standard errors in terms of posterior mean and standard deviation of the related parameters for all the three models. It may be noted that the model precision (\emph{i.e.} $\frac{1}{\sigma}$) is high for the ARSPI model indicating higher accuracy of the parameter estimates. Further, $\mid \beta_2 \mid \leq 1$ across all accumulation windows. High values of the $\beta_2$ values confirms the presence of auto-regression in MTR series. The probability of zero rainfall windows also remains stationary as the corresponding auto-regression coefficient $\phi$ is bounded in magnitude by unity.  In other words, the probability of drought does not explode to one in the longer time horizon.
\par \textbf{ARSPI Computation:} The computation of ARSPI is carried out similarly to the conventional SPI. To achieve this, the cumulative distribution function (CDF) must be estimated at each observed MTR data point. This estimation is performed using the predictive density. Let $\Tilde{\boldsymbol{\theta}}^{(i)}$ represent the $i^{\text{th}}$ draw from the posterior distribution $\pi(\boldsymbol{\theta} \mid \boldsymbol{r})$, where $i = 1, 2, \ldots, 45000$. A predictive sample of size 45000 is then generated at a specific time point $t$ by drawing one sample from the density $f(r_t \mid \Tilde{\boldsymbol{\theta}}^{(i)})$ for each $i$. The empirical CDF of this sample is then used to estimate the CDF at time $t$.

\subsection{Comparative Analysis of SPI and ARSPI in the Colorado River Basin}
This study presents the ARSPI as a flexible alternative to traditional SPI, leveraging Hierarchical Bayesian structure to better estimate meteorological drought conditions. The foundational SPI was developed by \cite{mckee1993relationship} using monthly precipitation data from the Colorado River Basin between January 1889 and December 1991.  

\par For this analysis, we compare ARSPI with SPI using almost same historical precipitation dataset spanning 1188 months from January 1893 onward \citep{Col2024data}. Both indices are computed over 3-, 6-, 12-, and 24-month accumulation windows. Shorter time windows (3 and 6 months) are useful for assessing agricultural impacts, whereas longer windows (12 and 24 months) are more relevant for water resource management decisions. 

\par The first plot in Figure~\ref{fig : spi_comp} shows SPI-3 and ARSPI-3 values for the Colorado River Basin. Both indices exhibit considerable variability, with SPI-3 values ranging from -3.6447 to 2.9582, while ARSPI-3 spans a wider range of -3.9798 to 3.4658. The oscillatory nature of the time series reflects climatic fluctuations with frequent transitions between dry and wet periods. The broader range of ARSPI suggests its enhanced ability to capture precipitation anomalies compared to SPI.  

\par Similarly, the second plot of Figure~\ref{fig : spi_comp} illustrates SPI-6 and ARSPI-6. While SPI-6 ranges between -3.1774 and 2.9838, ARSPI-6 shows sharper peaks and troughs, with a range from -4.0752 to 3.8118, emphasizing extreme climatic conditions more effectively.  

\par The third plot highlights SPI-12, which exhibits smoother oscillations over a range of -2.8892 to 2.4220, capturing long-term trends and annual precipitation anomalies. In contrast, ARSPI-12 demonstrates a higher amplitude, ranging from -3.8567 to 3.9109, making it more sensitive to short-term anomalies within annual cycles.  

\par Finally, the fourth plot compares SPI-24 and ARSPI-24. SPI-24 ranges from -2.7586 to 2.2796, suitable for analyzing prolonged droughts. However, ARSPI-24 shows more pronounced peaks and troughs, with a range of -4.0752 to 3.9109, indicating greater sensitivity to short-term variations within the 24-month window.

\subsubsection*{Mismatch Points in SPI Comparison}
Discrepancies between ARSPI and SPI are identified when the indices classify drought conditions differently. Two primary types of mismatches are defined as follows. \emph{Type 1} mismatch occurs ARSPI indicates wet conditions \(( ARSPI > 0 )\), while SPI suggests severe or extreme drought \(( SPI < -1.50 )\). \emph{Type 2} mismatch is defined when ARSPI classifies conditions as severe or extreme drought \(( ARSPI < -1.50 )\), whereas SPI indicates wet conditions \(( SPI > 0 )\).

In the present data set total 4.42\% is the mismatch rate, of which 4.59\% is of type 1 (4.65\%,3.99\%,5.14\%,4.61\% for 3,6,12,24 month accumulation windows) and 4.25\% is of type 2 (4.23\%,4.58\%,4.54\%,3.65\% for 3,6,12,24 month accumulation windows).
Figure~\ref{fig : mismatch_rainfall} illustrates these mismatches for accumulation windows of 3, 6, 12, and 24 months. Type 1 mismatches are shown by the blue dashed line, and Type 2 by the red solid line.\\
\textbf{3-Month Accumulation Window}:
Clusters of mismatches are observed in the early 1900s, 1940s, and 1960s. Type 2 events show higher rainfall variability, while Type 1 captures moderate to low rainfall values.\\
\textbf{6-Month Accumulation Window}:
Mismatches occur irregularly, with notable clustering in the 1920s and 1960s. Type 2 events are more frequent and associated with higher rainfall magnitudes.\\
\textbf{12-Month Accumulation Window}:
The 12-month window shows a dominance of Type 2 mismatches, often linked to significant rainfall anomalies, particularly in the early and mid-20th century.\\
\textbf{24-Month Accumulation Window}:
The 24-month window also demonstrates a dominance of Type 2 mismatches, often associated with significant rainfall anomalies, particularly in the early and mid-20th century.

\section{Conclusion} \label{sec : conclusion}

In this study, we introduce the AutoRegressive Standardized Precipitation Index (ARSPI), a novel drought index that incorporates the auto-correlated nature of precipitation series. By capturing evolving precipitation patterns under dynamic environmental conditions, ARSPI addresses limitations in traditional drought assessment methods. 

The key contribution of this work is the construction of ARSPI, and a comparative analysis of ARSPI and the conventional SPI across the Colorado River Basin. Utilizing Bayesian techniques, ARSPI provides a more flexible and adaptive framework for quantifying precipitation anomalies. Our evaluation across multiple accumulation windows (3, 6, 12, and 24 months) demonstrates that ARSPI outperforms SPI in identifying extreme drought and wet conditions. The broader range and pronounced oscillations observed in ARSPI, characterized by sharper peaks and troughs, enhance its sensitivity to climatic extremes, offering a more refined depiction of drought variability. 

Mismatch analyses between ARSPI and SPI reveal significant discrepancies in drought classification, particularly under extreme conditions. These differences highlight the nuanced response of each index to precipitation variability. The temporal clustering of mismatches underscores the heightened sensitivity of ARSPI to historical climatic patterns, consistently capturing extreme rainfall anomalies with greater accuracy.



ARSPI’s flexibility, enhanced sensitivity to precipitation extremes, and ability to capture nuanced drought characteristics make it a valuable tool for hydrological and climatic studies. Future research should explore its application in other regions and evaluate its integration into decision-making frameworks for drought mitigation and water resource management.

\section*{Appendix: Tables and Figures}
\subsection*{Tables}\label{Tables}
\begin{table}[ht]
\caption{Drought Classification using ARSPI} 
\centering
\begin{tabular}{|r|r|r|r|r|}
  \hline
  ARSPI & Category & Probability \\ 
  \hline
   ARSPI $\geq\;2.00$ & Extreme Wet & 0.023 \\   
  \hline
  $1.50 <$ ARSPI $< 1.99$ & Severe Wet & 0.044 \\ 
  \hline
  $1.00 <$ ARSPI $< 1.49$ & Moderate Wet & 0.092 \\ 
  \hline
  $0.00 <$ ARSPI $< 0.99$ & Mild Wet & 0.341 \\
  \hline
  $-0.99 <$ ARSPI $< 0.00$ & Mild Drought & 0.341 \\
  \hline
  $-1.49 <$ ARSPI $< -1.00$ & Moderate Drought & 0.092 \\
  \hline
 $-1.99 <$ ARSPI $< -1.50$ & Severe Drought & 0.044  \\
  \hline
  ARSPI $\leq\;-2.00$ & Extreme Drought & 0.023 \\
  \hline
\end{tabular}
\label{tab : SPItable}
\end{table}

\begin{table}[ht]
\caption{DIC values for the ARSPI model} 
\centering
\begin{tabular}{|r|r|}
  \hline
  Time & \emph{ARSPI Model} \\ 
  \hline
   3 MTR & 4632.639 \\   
  \hline
  6 MTR & 4526.911 \\ 
  \hline
  12 MTR & 3726.612 \\ 
  \hline
  24 MTR & 3636.744 \\
  \hline
\end{tabular}
\label{tab : DICtable}
\end{table}

\begin{table}[h]
\centering
\scriptsize
\caption{Parameter estimates for different Models.}
\begin{tabular}{|c|c|c|c|c|c|}
\hline
\textbf{Model} & \textbf{Parameters} & \textbf{3 MTR} & \textbf{6 MTR} & \textbf{12 MTR} & \textbf{24 MTR} \\ \hline
\multirow{5}{*}{ARSPI Model} & $\beta_1$ & 0.3756(0.0299) & 0.2465(0.0291) & 0.1388(0.0245) & 0.0885(0.0205) \\ \cline{2-6} 
                         & $\beta_2$ & 0.7220(0.0200) & 0.8854(0.0133) & 0.9523(0.0084) & 0.9755(0.0057) \\ \cline{2-6}
                         & $\sigma$ & 0.4411(0.0091) & 0.1916(0.0039) & 0.0650(0.0013) & 0.0313(0.0006) \\ \cline{2-6} 
                         & $\alpha$ & -6.8149(0.7851) & -6.8025(0.7835) & -6.8023(0.7827) & -6.7914(0.7847) \\ \cline{2-6} 
                         & $\phi$ & 0.0026(0.5724) & 0.0076(0.5716) & 0.0012(0.5707) & 0.0012(0.5698) \\ \hline
\end{tabular}
\label{tab:parest}
\end{table}

\newpage

\subsection*{Figures}\label{Figures}

\begin{figure}[!ht]
\centering
\includegraphics[width=0.49\textwidth]{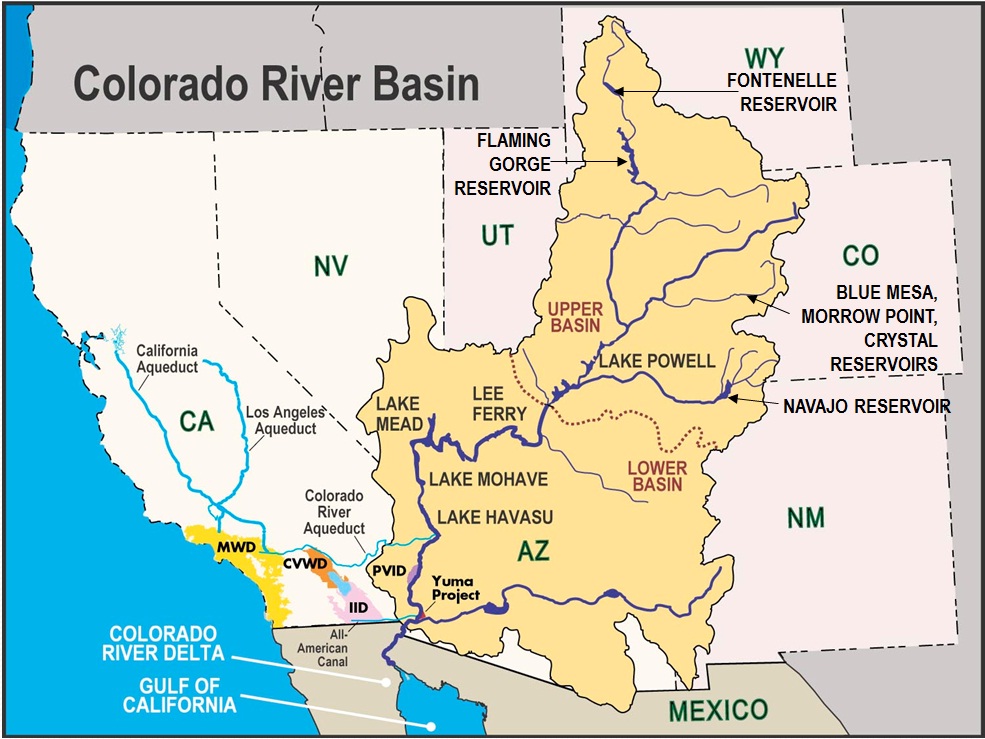}
\caption{Colorado River Basin.} 
\label{fig : colbasin}
\end{figure}

\begin{figure}[!ht]
\centering
\includegraphics[width=0.99\textwidth]{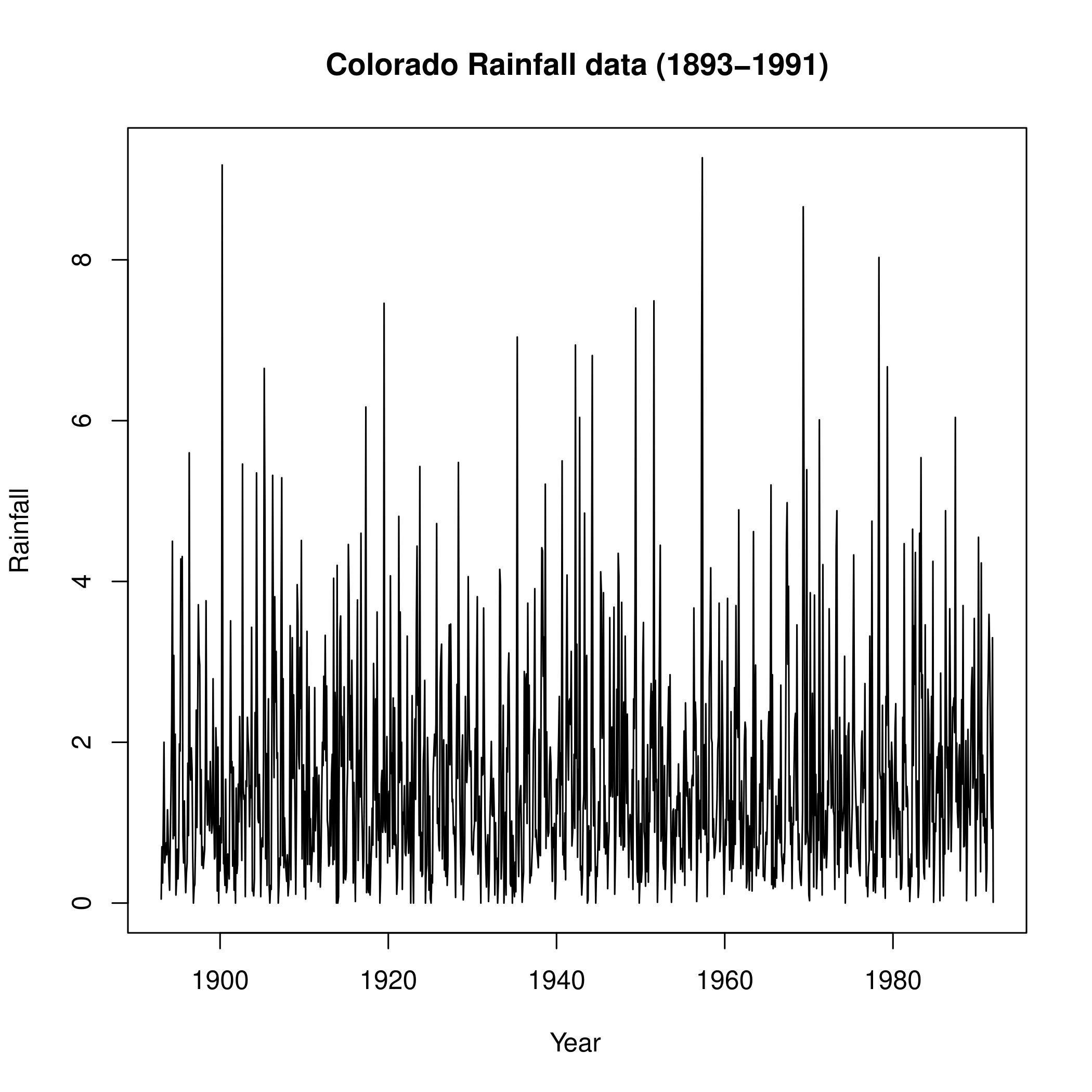}
\caption{Rainfall of Colorado River Basin.} 
\label{fig : raincol}
\end{figure}

\begin{figure}[!ht]
\centering
{\includegraphics[width=0.49\textwidth]{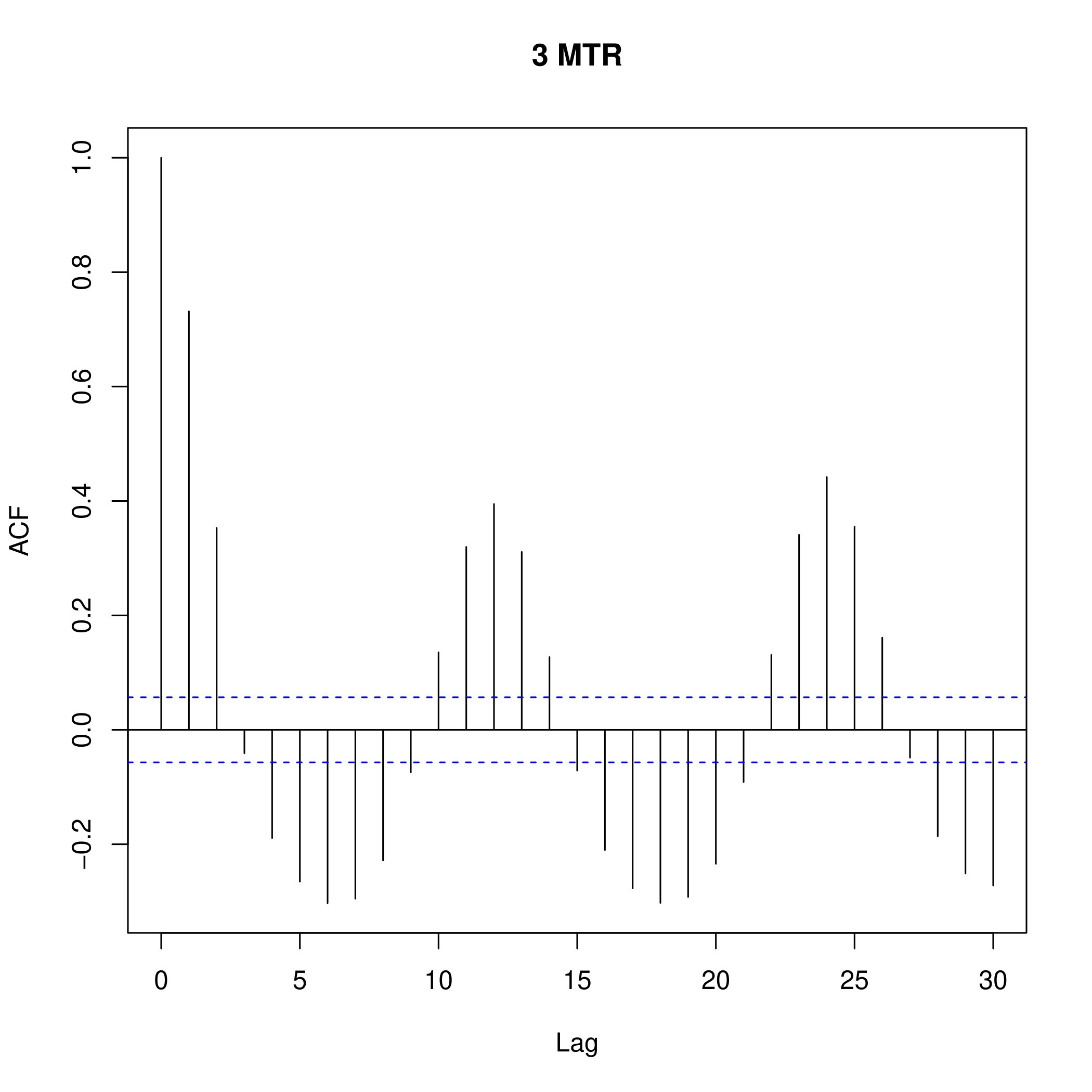}\label{fig:3mtr_acf}}
\hfill
{\includegraphics[width=0.49\textwidth]{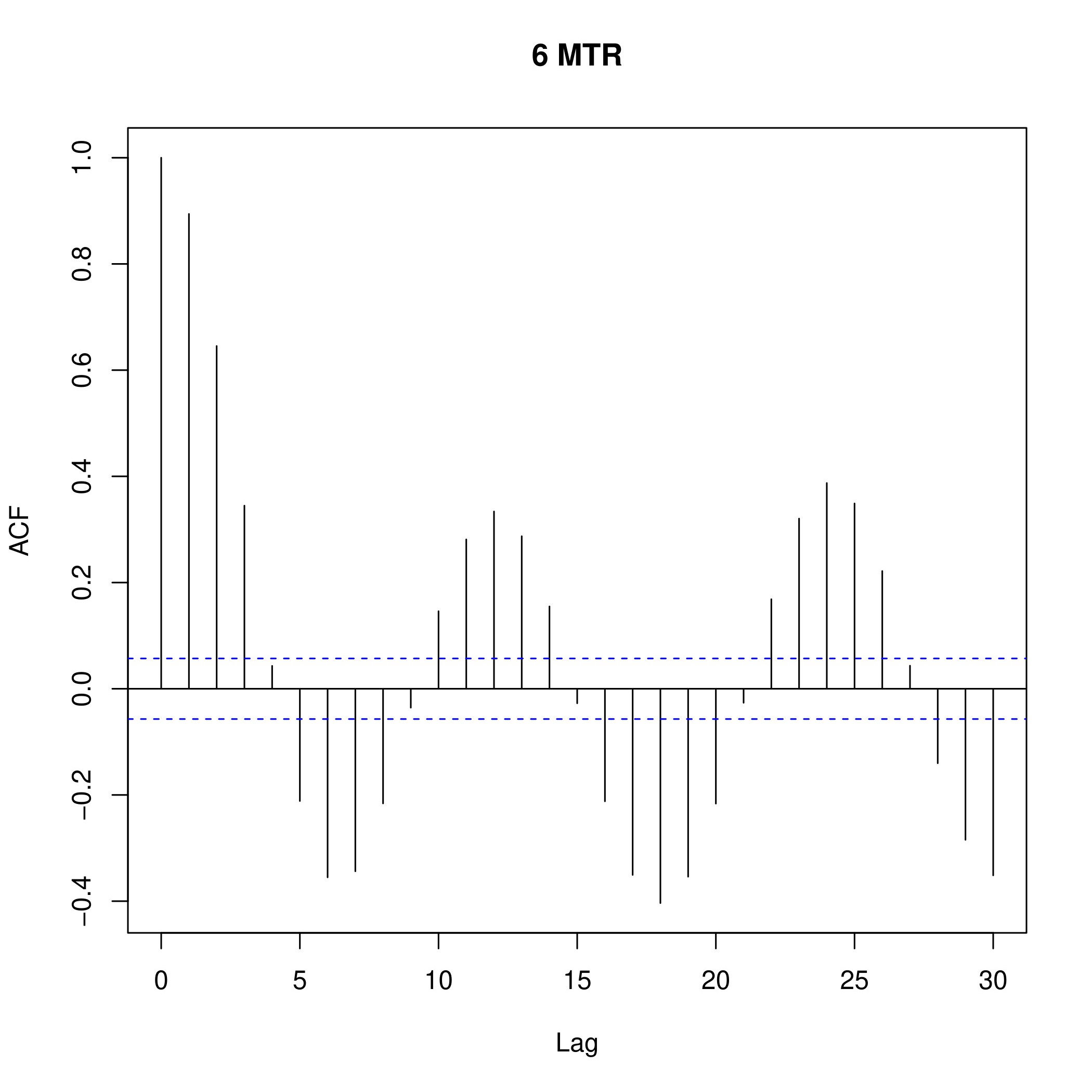}\label{fig:6mtr_acf}}\\ 
{\includegraphics[width=0.49\textwidth]{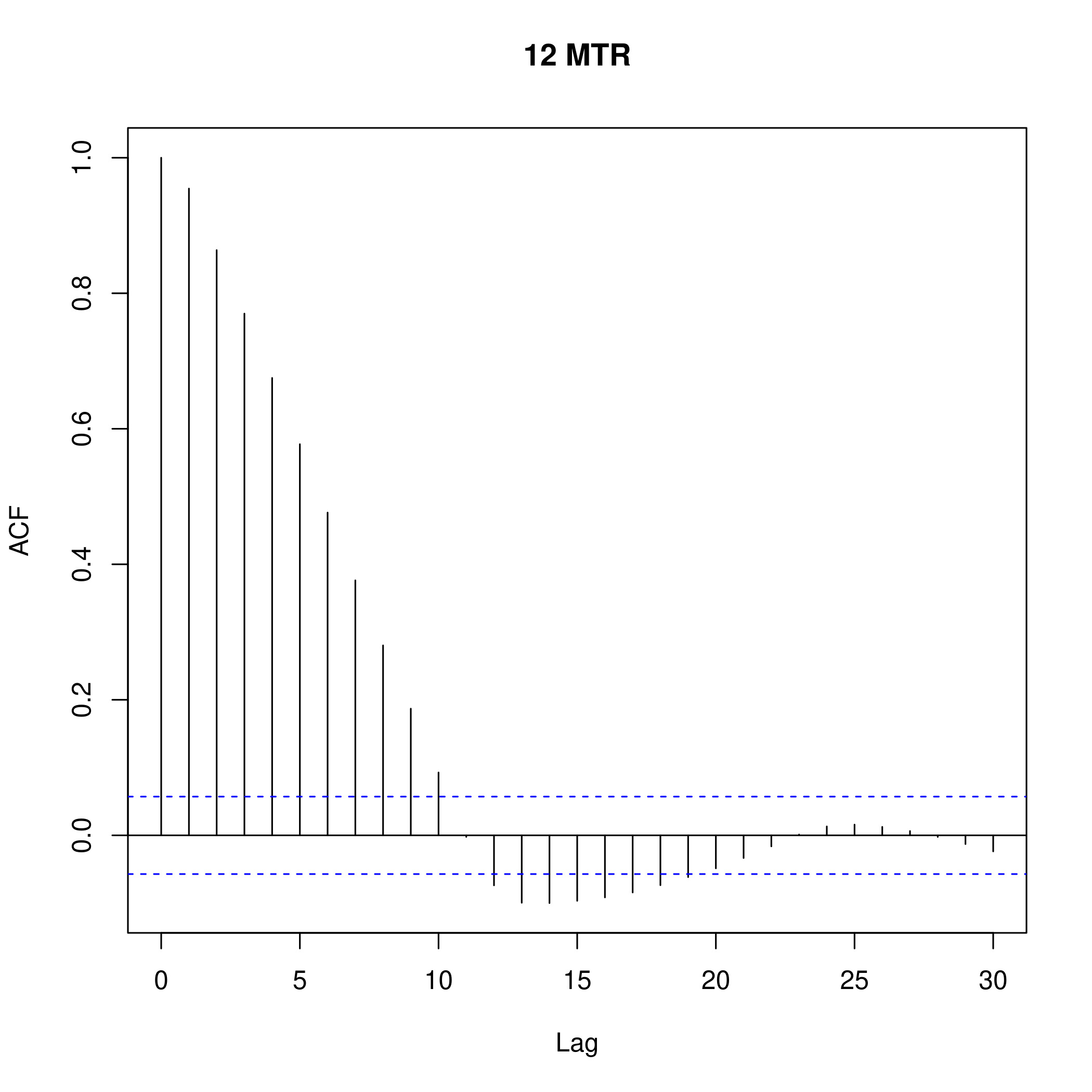}\label{fig:12mtr_acf}}
\hfill
{\includegraphics[width=0.49\textwidth]{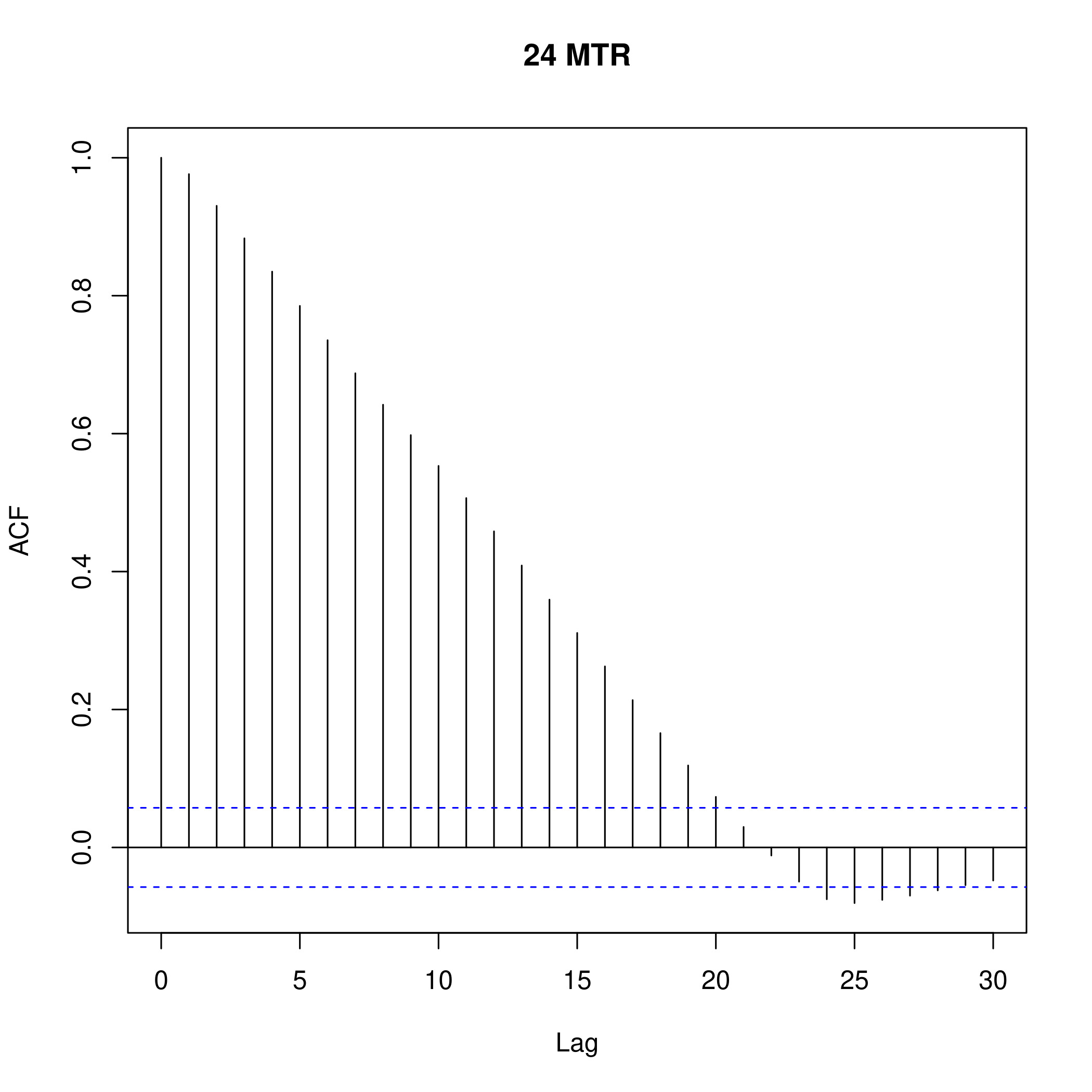}\label{fig:24mtr_acf}}
\caption{ACF plots of across different accumulation windows.} 
\label{fig : acfplt}
\end{figure}

\begin{figure}[!ht]
\centering
{\includegraphics[width=0.49\textwidth]{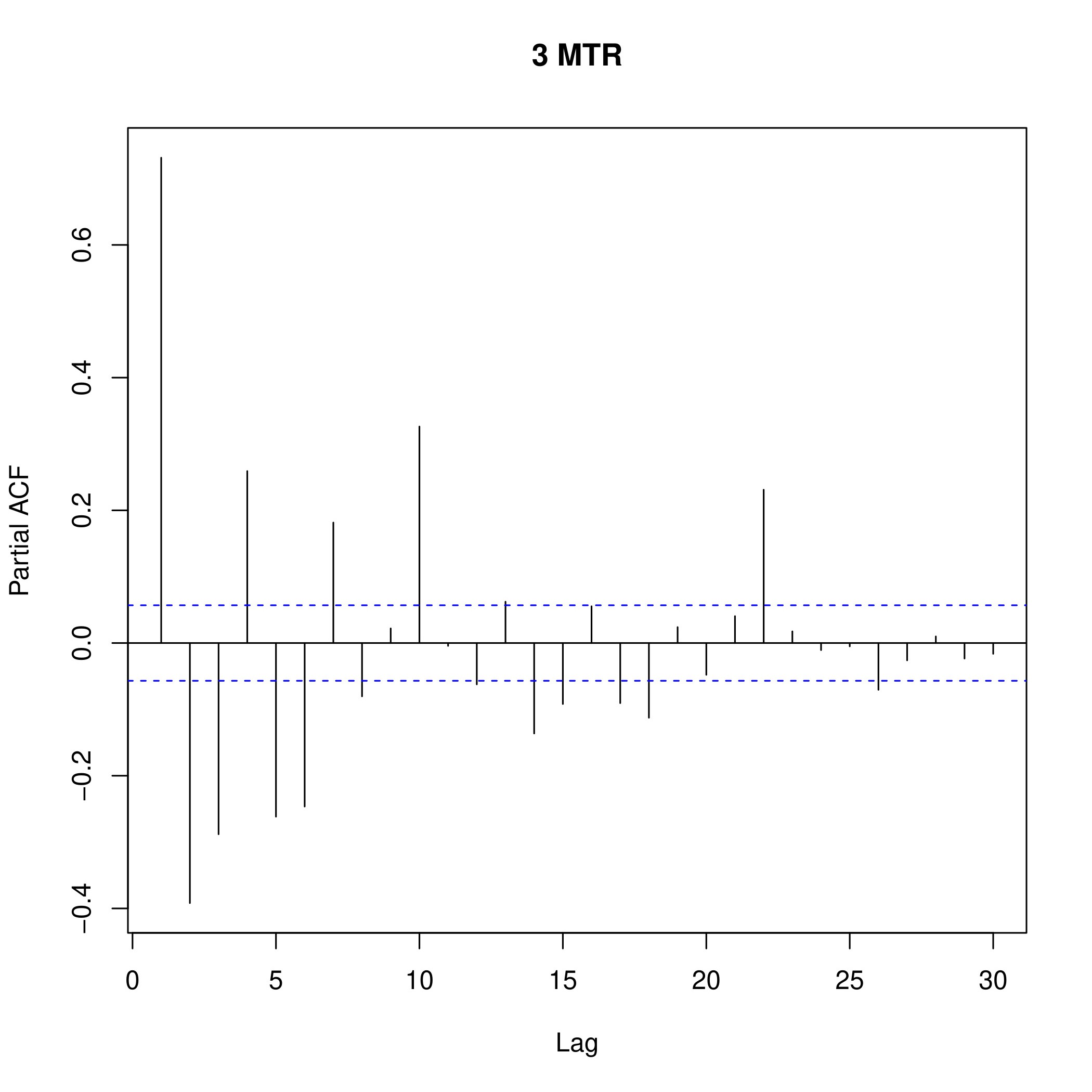}\label{fig:3mtr_pacf}}
\hfill
{\includegraphics[width=0.49\textwidth]{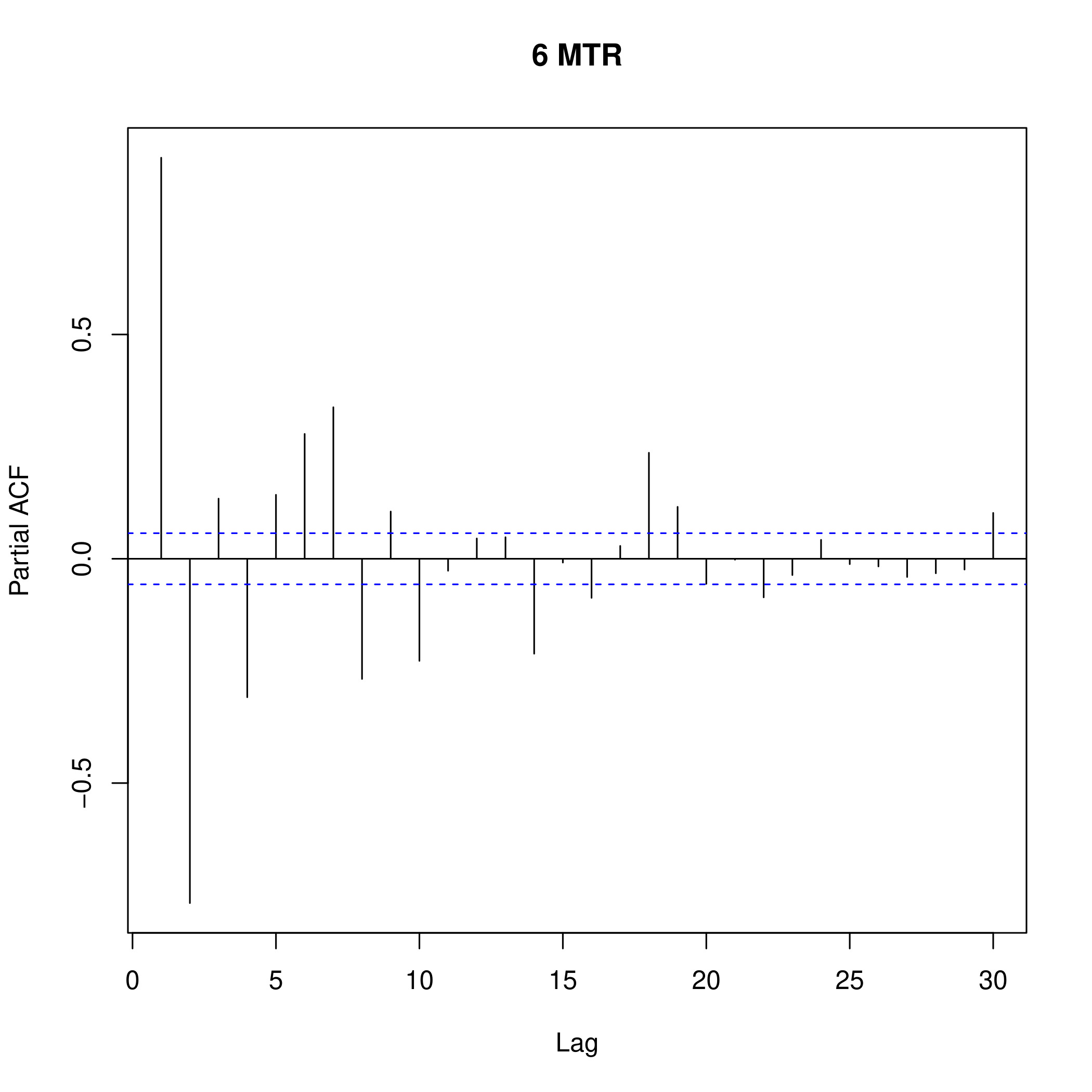}\label{fig:6mtr_pacf}}\\ 
{\includegraphics[width=0.49\textwidth]{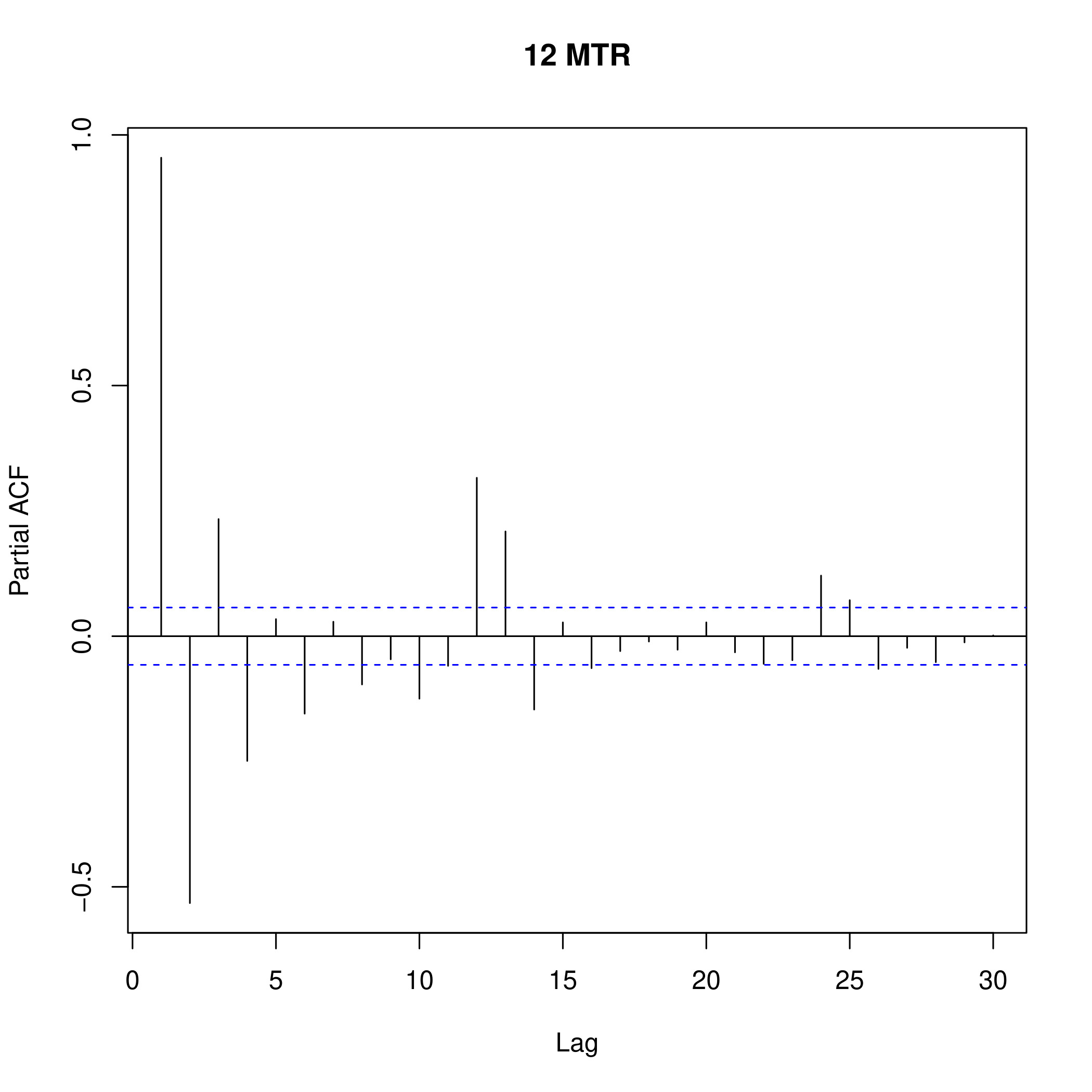}\label{fig:12mtr_pacf}}
\hfill
{\includegraphics[width=0.49\textwidth]{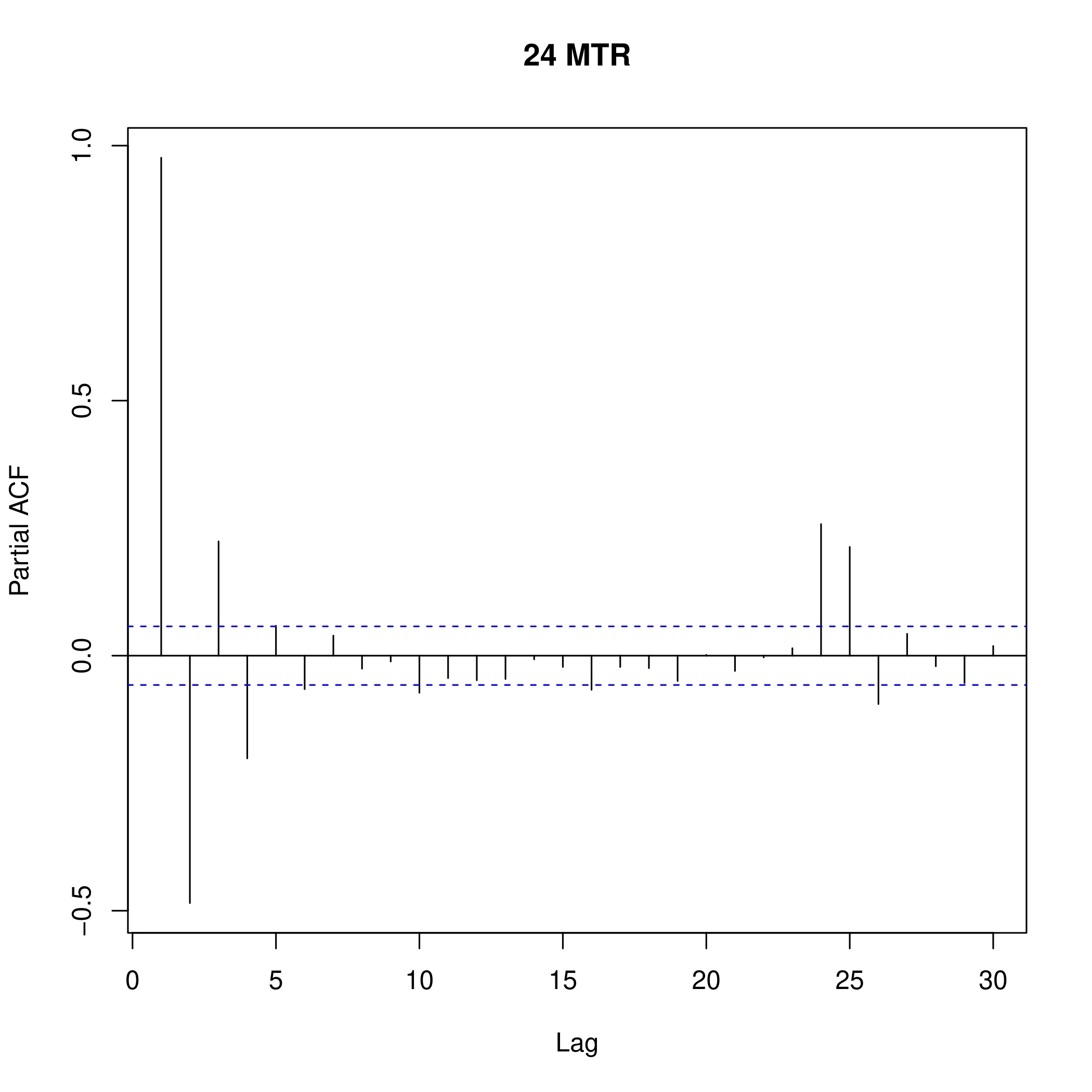}\label{fig:24mtr_pacf}}
\caption{Partial ACF plots across different accumulation windows.} 
\label{fig : pacfplt}
\end{figure}

\begin{figure}[!ht]
\centering
{\includegraphics[width=0.99\textwidth]{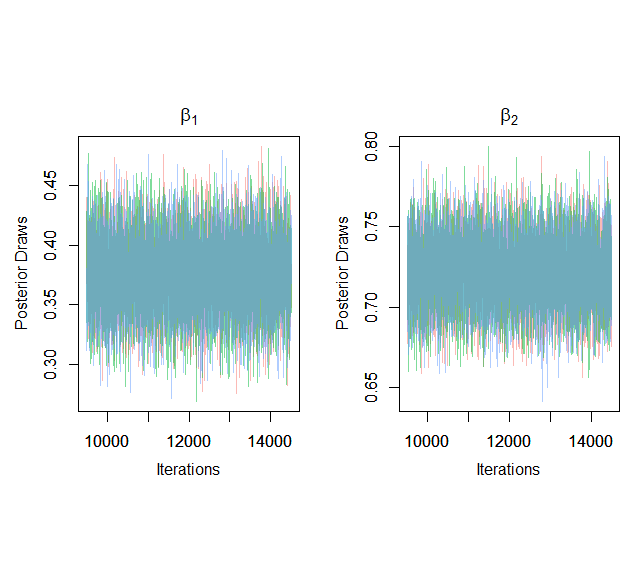}\label{fig:spiar1_hbpt3beta_tr}}\\
{\includegraphics[width=0.32\textwidth]{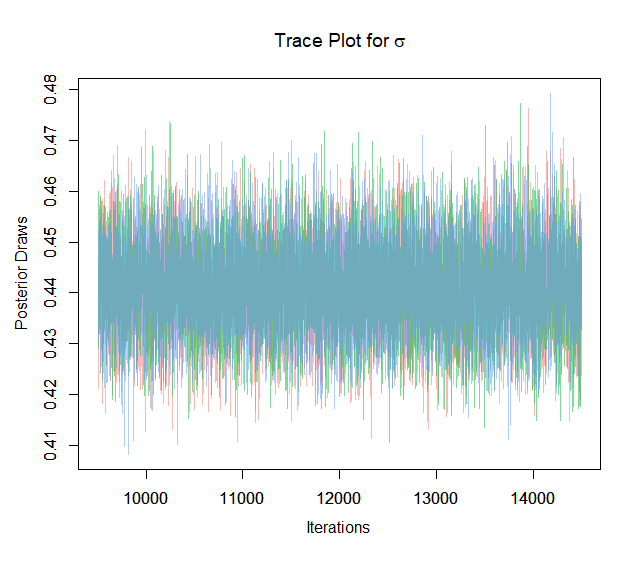}\label{fig:spiar1_hbpt3sigma_tr}} 
\hfill
{\includegraphics[width=0.32\textwidth]{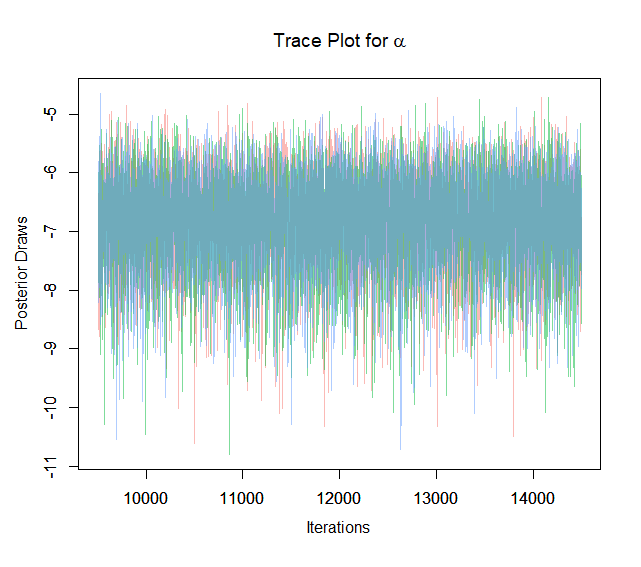}\label{fig:spiar1_hbpt3alpha_tr}}
\hfill
{\includegraphics[width=0.32\textwidth]{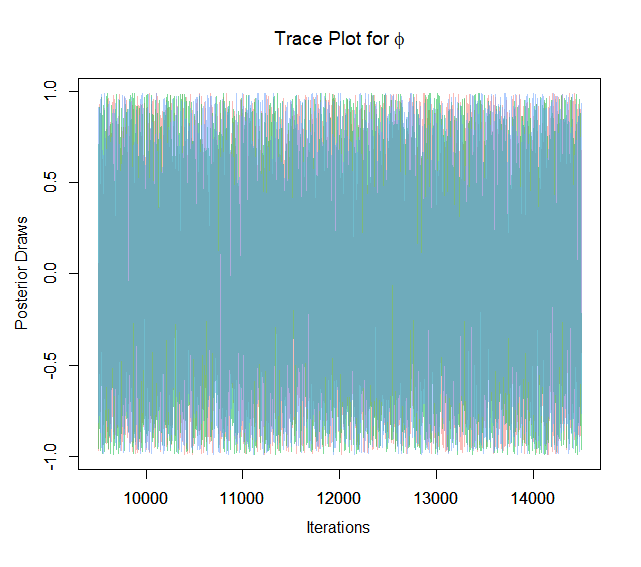}\label{fig:spiar1_hbpt3phi_tr}}
\caption{Trace plots for parameters for \emph{Model III} (3 month accumulation window).} 
\label{fig : spiar1_hbpt3_tr}
\end{figure}

\begin{figure}[!ht]
\centering
{\includegraphics[width=0.99\textwidth]{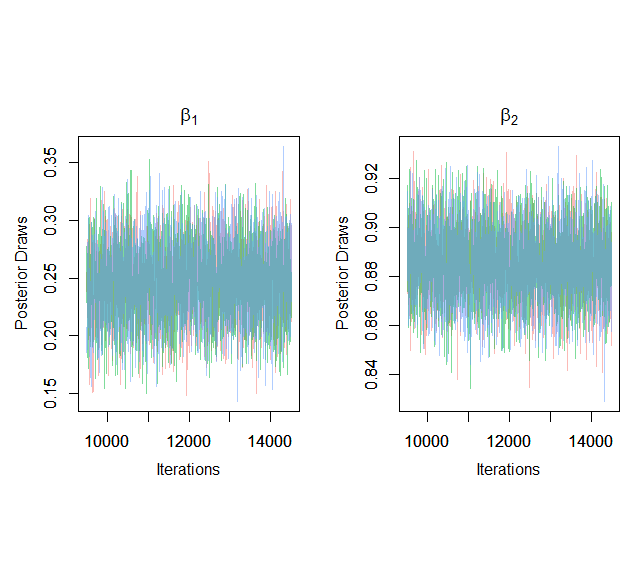}\label{fig:spiar1_hbpt6beta_tr}}\\
{\includegraphics[width=0.32\textwidth]{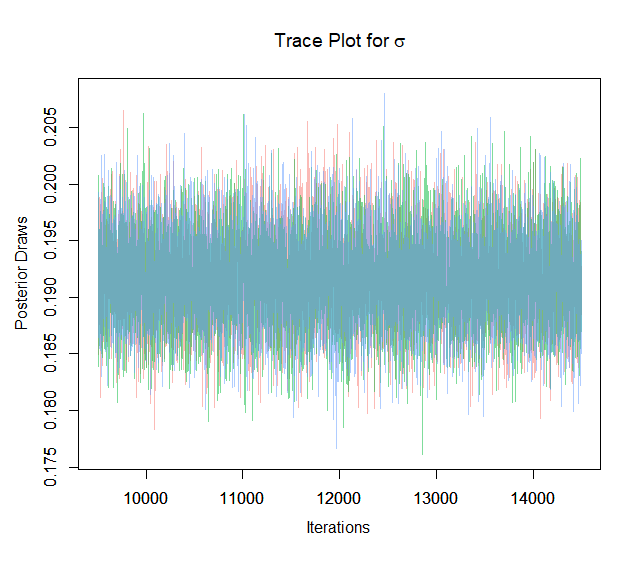}\label{fig:spiar1_hbpt6sigma_tr}} 
\hfill
{\includegraphics[width=0.32\textwidth]{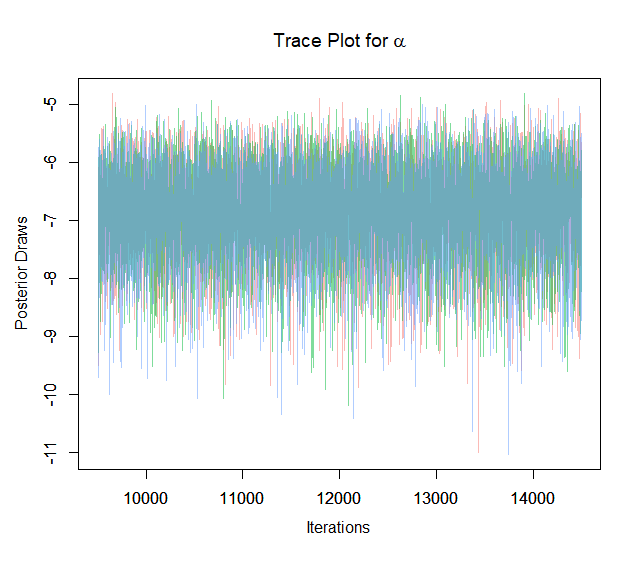}\label{fig:spiar1_hbpt6alpha_tr}}
\hfill
{\includegraphics[width=0.32\textwidth]{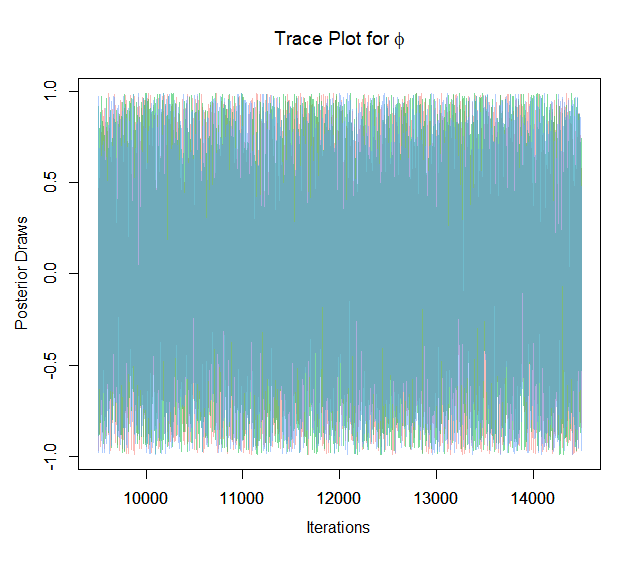}\label{fig:spiar1_hbpt6phi_tr}}
\caption{Trace plots for parameters for \emph{Model III} (6 month accumulation window).} 
\label{fig : spiar1_hbpt6_tr}
\end{figure}

\begin{figure}[!ht]
\centering
{\includegraphics[width=0.99\textwidth]{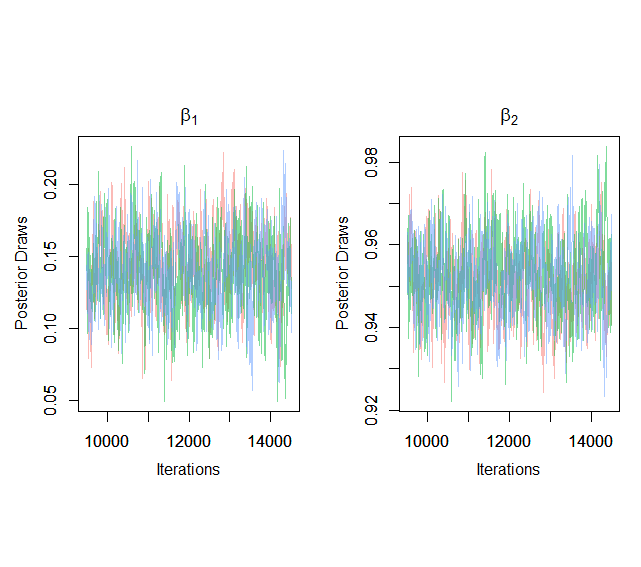}\label{fig:spiar1_hbpt12beta_tr}}\\
{\includegraphics[width=0.32\textwidth]{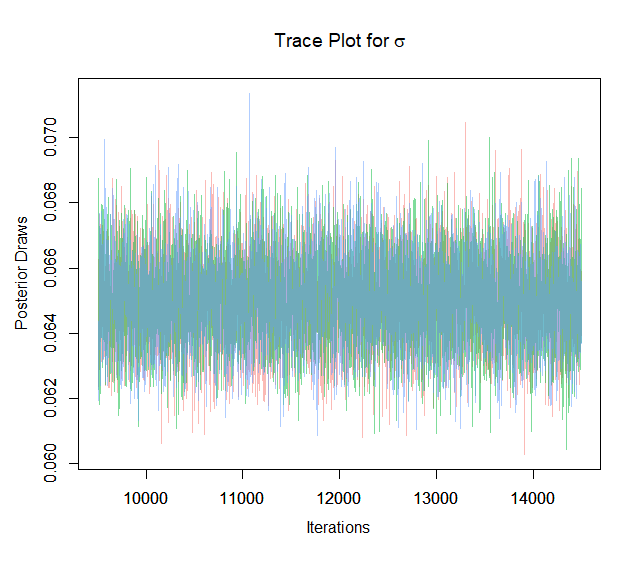}\label{fig:spiar1_hbpt12sigma_tr}} 
\hfill
{\includegraphics[width=0.32\textwidth]{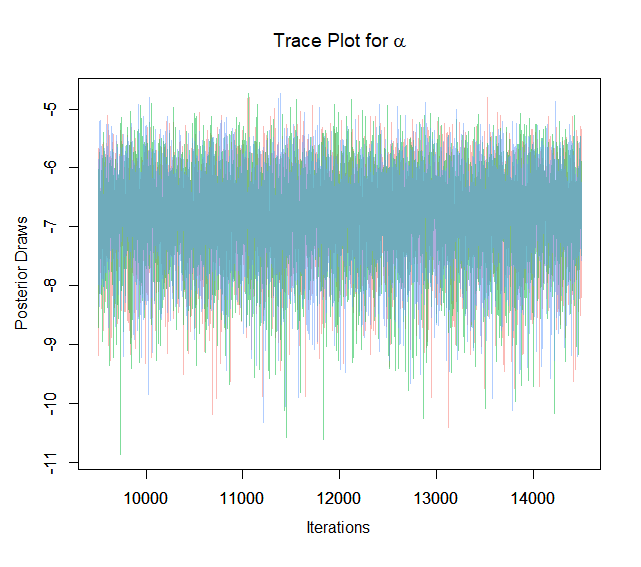}\label{fig:spiar1_hbpt12alpha_tr}}
\hfill
{\includegraphics[width=0.32\textwidth]{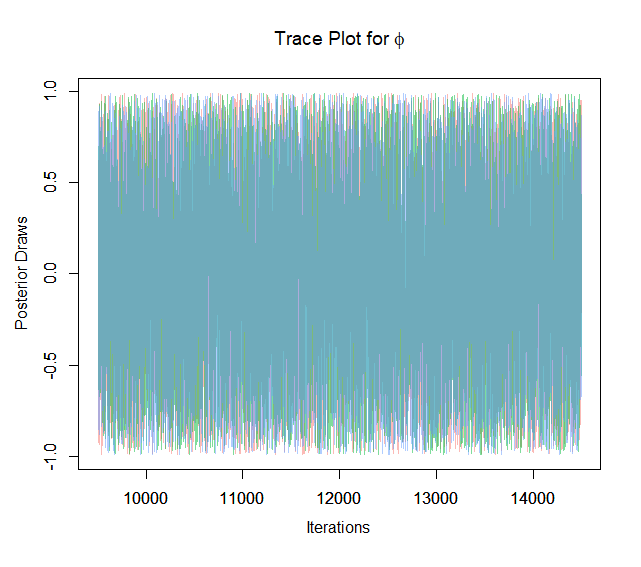}\label{fig:spiar1_hbpt12phi_tr}}
\caption{Trace plots for parameters for \emph{Model III} (12 month accumulation window).} 
\label{fig : spiar1_hbpt12_tr}
\end{figure}

\begin{figure}[!ht]
\centering
{\includegraphics[width=0.99\textwidth]{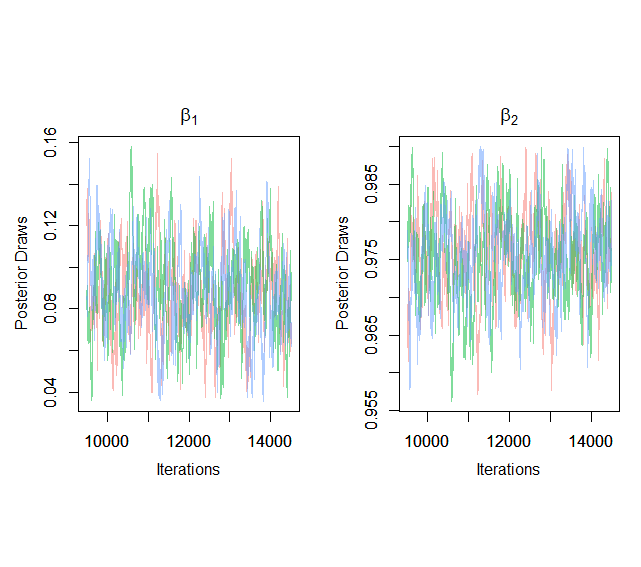}\label{fig:spiar1_hbpt24beta_tr}}\\
{\includegraphics[width=0.32\textwidth]{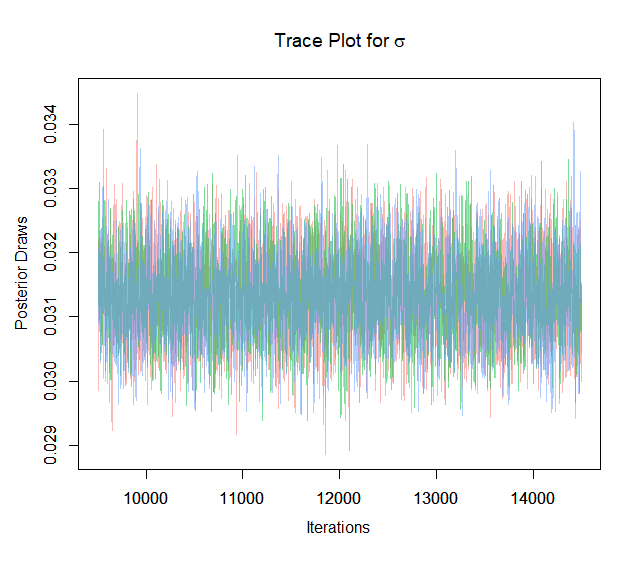}\label{fig:spiar1_hbpt24sigma_tr}} 
\hfill
{\includegraphics[width=0.32\textwidth]{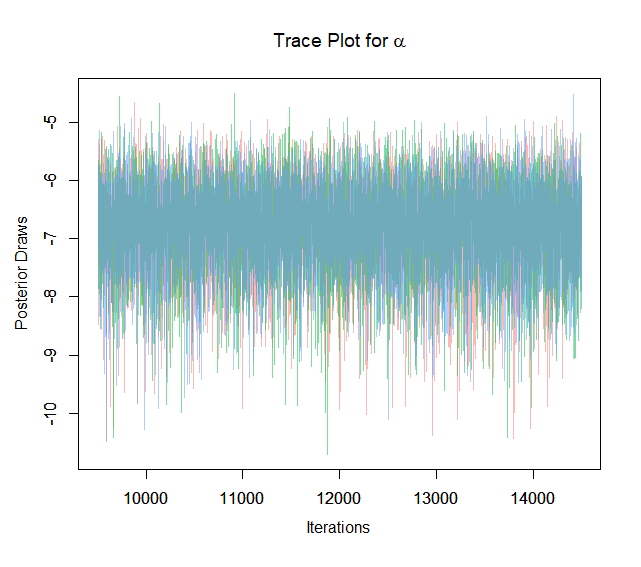}\label{fig:spiar1_hbpt24alpha_tr}}
\hfill
{\includegraphics[width=0.32\textwidth]{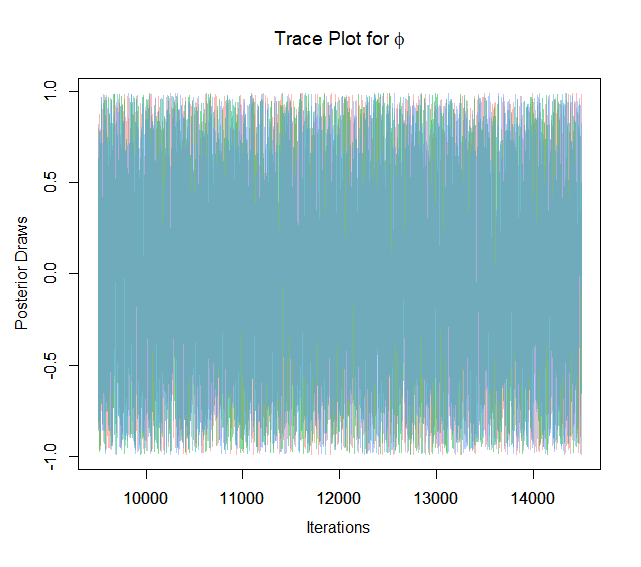}\label{fig:spiar1_hbpt24phi_tr}}
\caption{Trace plots for parameters for \emph{Model III} (24 month accumulation window).} 
\label{fig : spiar1_hbpt24_tr}
\end{figure}

\begin{figure}[!ht]
\centering
{\includegraphics[width=0.49\textwidth]{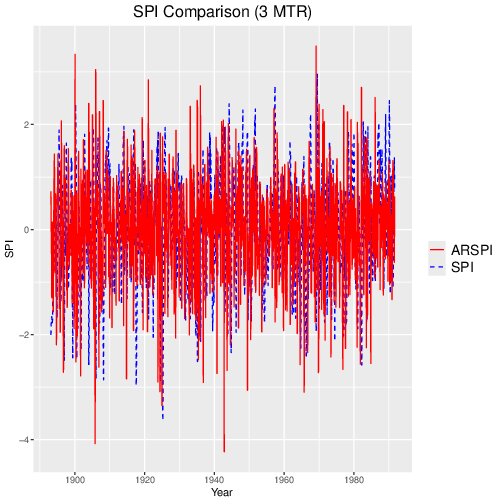}\label{fig : spi3_comp}}
\hfill
{\includegraphics[width=0.49\textwidth]{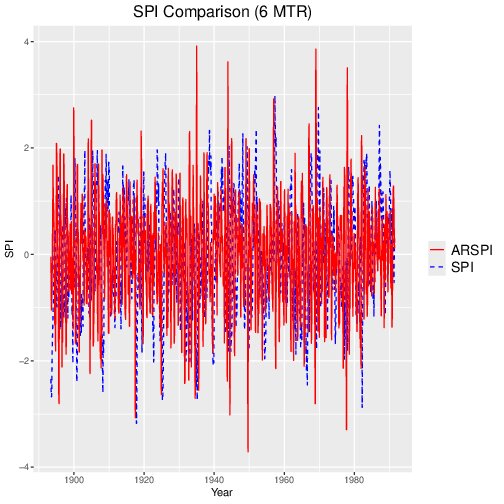}\label{fig : spi6_comp}}\\ 
{\includegraphics[width=0.49\textwidth]{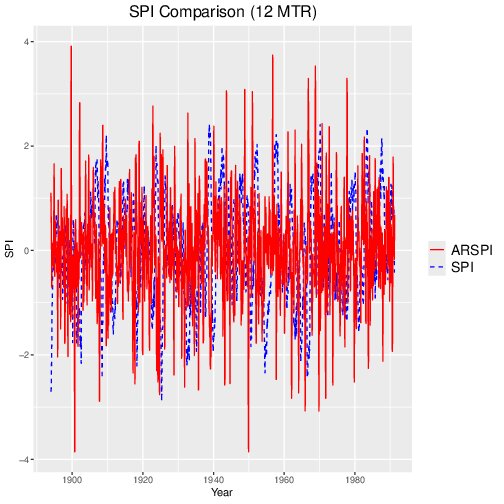}\label{fig : spi12_comp}}
\hfill
{\includegraphics[width=0.49\textwidth]{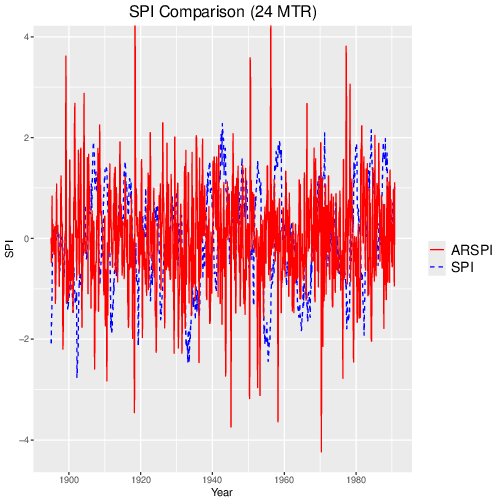}\label{fig : spi24_comp}}
\caption{Comparison of SPI and ARSPI for different accumulation windows (3,6,12,24 months).} 
\label{fig : spi_comp}
\end{figure}

\begin{figure}[!ht]
\centering
{\includegraphics[width=0.49\textwidth]{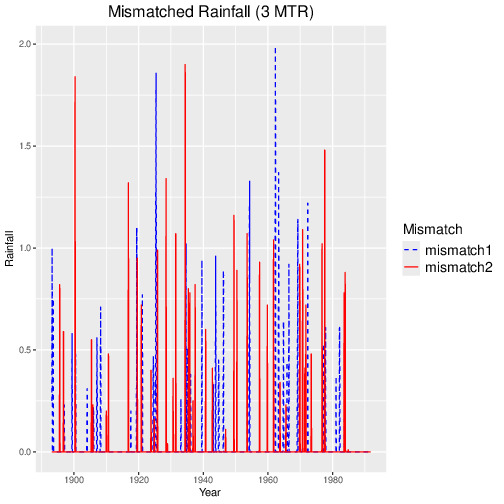}\label{fig : mismatch_3}}
\hfill
{\includegraphics[width=0.49\textwidth]{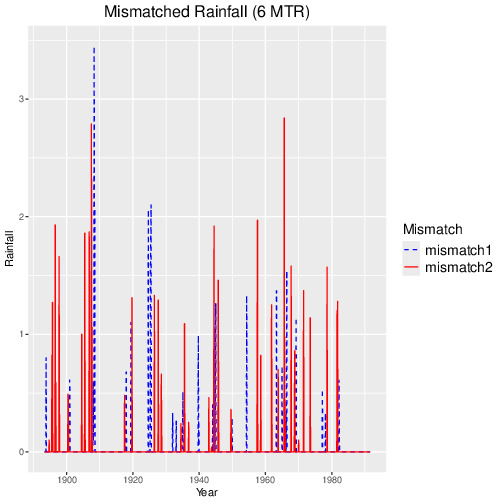}\label{fig : mismatch_6}}\\ 
{\includegraphics[width=0.49\textwidth]{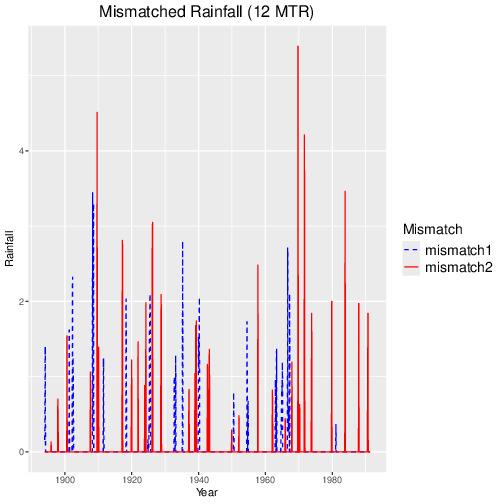}\label{fig : mismatch_12}}
\hfill
{\includegraphics[width=0.49\textwidth]{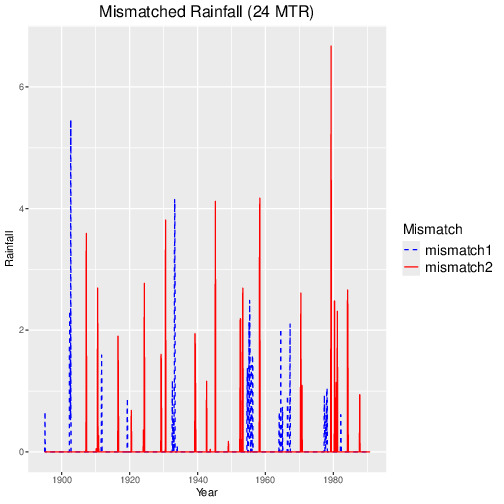}\label{fig : mismatch_24}}
\caption{Plots representing the mismatched rainfall for different accumulation windows (3,6,12,24 months).} 
\label{fig : mismatch_rainfall}
\end{figure}






\end{document}